\documentclass[a4paper,fleqn]{cas-sc}

\usepackage[numbers]{natbib}
\usepackage{multirow} 
\usepackage{array}    

\def\tsc#1{\csdef{#1}{\textsc{\lowercase{#1}}\xspace}}
\tsc{WGM}
\tsc{QE}
\tsc{EP}
\tsc{PMS}
\tsc{BEC}
\tsc{DE}

\begin{document}
\let\WriteBookmarks\relax
\def\floatpagepagefraction{1}
\def\textpagefraction{.001}
\shorttitle{A High-Performance External Validity Index for Clustering with a Large Number of Clusters}
\shortauthors{M.Y. Karbasian et al.}

\title [mode = title]{A High-Performance External Validity Index for Clustering with a Large Number of Clusters}

\author[1]{Mohammad Yasin Karbasian}[type=editor,
                        auid=000,bioid=1,
                        orcid=0009-0009-1255-2340]
\cormark[1]
\fnmark[1]
\ead{m.karbasian@ec.iut.ac.ir}


\affiliation[1]{organization={Department of Electrical and Computers Engineering, Isfahan University of Technology},
                city={Isfahan},
                state={Isfahan},
                country={Iran}}

\author[2]{Ramin Javadi}[]



\affiliation[2]{organization={Department of Mathematical Sciences, Isfahan University of Technology},
                city={Isfahan},
                state={Isfahan},
                country={Iran}}



\cortext[cor1]{Corresponding author}

\begin{abstract}
This paper introduces the Stable Matching Based Pairing (SMBP) algorithm, a high-performance external validity index for clustering evaluation in large-scale datasets with a large number of clusters. SMBP leverages the stable matching framework to pair clusters across different clustering methods, significantly reducing computational complexity to $O(N^2)$, compared to traditional Maximum Weighted Matching (MWM) with $O(N^3)$ complexity. Through comprehensive evaluations on real-world and synthetic datasets, SMBP demonstrates comparable accuracy to MWM and superior computational efficiency. It is particularly effective for balanced, unbalanced, and large-scale datasets with a large number of clusters, making it a scalable and practical solution for modern clustering tasks. Additionally, SMBP is easily implementable within machine learning frameworks like PyTorch and TensorFlow, offering a robust tool for big data applications. The algorithm is validated through extensive experiments, showcasing its potential as a powerful alternative to existing methods such as Maximum Match Measure (MMM) and Centroid Ratio (CR). 
\end{abstract}


\begin{highlights}
\item SMBP reduces clustering validation complexity from $O(N^3)$ to $O(N^2)$.

\item SMBP efficiently scales for large datasets with numerous clusters.

\item SMBP achieves similar accuracy to traditional methods with faster performance.

\item SMBP integrates smoothly with PyTorch and TensorFlow for machine learning.

\item SMBP works well with both balanced and unbalanced clustering datasets.
\end{highlights}

\begin{keywords}
Stable Matching \sep External Validity Index \sep Clustering Evaluation \sep Big Data \sep Maximum Weighted Matching
\end{keywords}

\maketitle

\section{Introduction}
Clustering is a fundamental technique in data analysis, used to partition datasets into meaningful groups or clusters based on similarities between data points \cite{xu2005survey}. It finds applications in diverse fields, including data mining \cite{bijuraj2013clustering}, bioinformatics \cite{masood2015clustering}, computer vision \cite{jolion1991robust}, and market segmentation \cite{ramasubbareddy2020comparative}. However, one of the challenges in clustering is determining how well a particular partition fits the data, especially when multiple clustering algorithms or parameter settings are applied \cite{wu2009adapting, wu2009external, vinh2010information}. This is where external validity indexes come into play, providing a mechanism for comparing different clustering results by evaluating their consistency with ground truth labels or other clustering solutions \cite{dom2012information, franti2014centroid}. External validity indexes help assess the quality and stability of clustering methods, aiding in tasks such as determining the number of clusters \cite{law2003cluster, falasconi2010stability, pascual2010cluster}. 

External validity indexes can be broadly classified into three categories: pair-counting, information-theoretic, and set-matching. Pair-counting measures, including the Rand index \cite{rand}, Adjusted Rand index \cite{Hubert1985}, and Jaccard coefficient \cite{jaccardpaul}, focus on comparing object pairs from different clusterings. These methods evaluate whether object pairs placed in the same cluster in one partition are similarly clustered in another, representing agreement between partitions. Information-theoretic measures like entropy \cite{taoli}, mutual information \cite{romano2014standardized}, and variation of information \cite{MEILA2007873} compare clusterings based on shared information. Since mutual information lacks an upper bound, normalization is typically applied for better interpretability. Set-matching measures such as the F-measure \cite{dalli2003adaptation}, and Van Dongen index \cite{van2000performance} pair similar clusters between partitions. 

Among the various methods used for clustering comparison, the Maximum Weighted Matching (MWM)  algorithm offers a promising approach \cite{rezaei2016set}. MWM can be adapted as an external validity index by treating clustering solutions as sets of clusters that need to be optimally paired. In this setting, every pair of clusters from different partitions is assigned a weight that represents a measure of similarity between clusters and the goal is to find a pairing (a one-to-one correspondence) between clusters while maximizing the sum of the weights of selected pairs. While this approach is effective in identifying optimal pairings, MWM operates in $O(N^3)$ time, making it computationally prohibitive for datasets with a large number of clusters. To address this issue, various approximation techniques have been proposed to reduce the computational complexity of cluster comparison methods, enabling their use on datasets with a larger number of clusters without sacrificing too much accuracy. 

The field of approximation methods for maximum weighted matching has seen significant advancements over the years, with each study building upon the work of its predecessors to achieve better results. Gabow and Tarjan made foundational contributions in 1988 \cite{gabow1988algorithms} and 1991 \cite{gabow1991faster}, laying the groundwork for efficient matching algorithms. Later, Preis \cite{preis1999linear} in 1999 introduced a notable linear-time \( \frac{1}{2} \)-approximation algorithm, which marked a significant improvement in the practical application of these methods. Drake and Hougardy \cite{drake2003simple} in 2003 further refined these approaches by proposing more effective algorithms that offered better approximation guarantees. Pettie and Sanders \cite{pettie2004simpler} in 2004 continued this trend, enhancing the run time of approximation methods.

In 2010, Duan and Pettie \cite{duan2010approximating}, as well as Hanke and Hougardy \cite{hanke2010new}, made substantial contributions by developing algorithms that improved upon the previous studies, offering more efficient solutions with better approximation ratios. The most significant breakthrough came in 2014 when Duan and Pettie \cite{duan2014linear} introduced a novel algorithm that bypassed the long-standing \( m \sqrt{n} \) barrier in approximation. Their algorithm could compute a \( (1 - \epsilon) \)-approximate maximum weight matching in \( O\left(m \epsilon^{-1} \log \epsilon^{-1}\right) \) time, achieving optimal linear time for any fixed \( \epsilon \). This new method is not only simpler than the best exact algorithms but also highly effective in applications where a negligible relative error is acceptable.

However, today’s datasets are becoming so large and complex that even these older approximation algorithms struggle to provide timely results, making them less practical for real-world applications where computational efficiency is critical. Additionally, we also need approaches that can be easily implemented within modern machine learning frameworks like PyTorch \cite{paszke2017automatic} or TensorFlow \cite{tensorflow2015-whitepaper}, especially for creating gradients. Unfortunately, many of the existing methods are not well-suited for integration into these frameworks, further limiting their practical utility.

In this context, we introduce the Stable Matching Based Pairing (SMBP) algorithm as an alternative to MWM for cluster pairing. SMBP leverages the concept of stable matching, where clusters from two different partitions are treated as entities with preferences. By constructing these preferences using the contingency matrix, SMBP performs the pairing in $O(N^2)$ time, significantly reducing the computational cost. Our method approximates MWM while maintaining high accuracy, making it particularly well-suited for large-scale clustering tasks.

In this paper, we demonstrate that SMBP not only outperforms other techniques in terms of computational efficiency, especially for big data applications but also retains a high level of accuracy in pairing clusters. Our experimental results show that SMBP achieves scalability without compromising the quality of clustering comparisons, positioning it as a robust and efficient alternative for external validity indexing in clustering analysis. Moreover, this approach is simple to program and can be easily implemented within modern machine learning frameworks such as PyTorch and TensorFlow, making it highly accessible for practical use in large-scale clustering tasks. Also, our method can compare two cluster groups with different numbers of clusters.

The field of clustering has seen significant advancements in recent years, particularly in the development of methods for comparing clustering results. One such method is the maximum match measure \cite{Wagner2007ComparingC}. This method evaluates the similarity between two clustering solutions by representing them as a bipartite graph, where clusters from the two solutions form the two sets of vertices. The edges between these sets are weighted by the similarity between corresponding clusters. The algorithm employs a greedy approach to iteratively pair clusters by selecting edges with the highest weights, maximizing the total matching score. However, because it is a greedy approach, it does not guarantee an optimal solution. While useful, the method has a computational complexity of $O(N^3)$, which can be demanding for datasets with a large number of clusters. Another advancement is the centroid ratio method, proposed by Zhao and Fränti \cite{Zhao}. This technique simplifies the evaluation of clustering by focusing on the nearest pairing of centroids between two clustering solutions. Also, this approach uses a greedy algorithm similar to maximum match measure to efficiently pair centroids based on distance.

The primary contributions of this paper are:

\begin{enumerate}
\item \textbf{SMBP Algorithm}: We introduce the SMBP algorithm, which leverages stable matching principles to efficiently compare clustering solutions.
\item \textbf{Comparative Evaluation}: We provide an extensive empirical evaluation of SMBP against other leading methods in terms of both accuracy and runtime efficiency.
\item \textbf{Scalability}: Our experiments demonstrate that SMBP scales well with datasets with a large number of clusters, making it a suitable choice for large-scale clustering tasks.
\end{enumerate}

The remainder of this paper is structured as follows: 
\begin{itemize}
    \item Section 2: Methods – This section details the methodologies used, including the contingency matrix, maximum weighted matching, stable matching, stable matching-based pairing, the process for generating synthetic datasets, and the experiments conducted.
 \item Section 3: Results – This section presents the experimental results, focusing on model runtime and accuracy.
 \item Section 4: Discussion – Here, we address potential misunderstandings regarding our approach and demonstrate its practical application.
 \item Section 5: Conclusion – The paper concludes with a summary of the key findings and contributions of this study.
\end{itemize}

\section{Methods}

\subsection{Contingency Matrix}

A \textit{contingency matrix} is a fundamental tool used in clustering analysis to evaluate the similarity between two different clustering results. It provides a representation of how the data points are distributed across the clusters of two different clustering methods. In this matrix, the entry $a_{ij}$ represents the number of data points that are placed in the $i$-th cluster of the first clustering result and the $j$-th cluster of the second clustering result.

Mathematically, if we have two clustering solutions:
\begin{itemize}
    \item First clustering result, denoted as $\mathcal{C}_1$, consists of $k_1$ clusters.
    \item Second clustering result, denoted as $\mathcal{C}_2$, consists of $k_2$ clusters.
\end{itemize}

The contingency matrix $M$ is of size $k_1 \times k_2$, where each entry $a_{ij}$ is computed as:
\[
a_{ij} = |\mathcal{C}_1^i \cap \mathcal{C}_2^j|
\]

Here, $\mathcal{C}_1^i$ denotes the set of data points assigned to the $i$-th cluster of the first clustering result, and $\mathcal{C}_2^j$ denotes the set of data points assigned to the $j$-th cluster of the second clustering result. Thus, $a_{ij}$ counts the number of common data points between the $i$-th cluster of the first clustering and the $j$-th cluster of the second clustering.

\subsection{Maximum Weighted Matching (MWM)}

Maximum Weighted Matching (MWM) is a classical problem in combinatorial optimization, where the objective is to find the optimal pairing between two sets of elements such that the total weight of the paired edges is maximized. In the context of a weighted bipartite graph, the weight of an edge represents the value associated with pairing two specific elements.

Formally, given a bipartite graph \(G = (U \cup V, E)\) with \(U\) and \(V\) as disjoint sets of vertices and \(E\) as the set of edges connecting vertices in \(U\) to vertices in \(V\), each edge \((u_i, v_j) \in E\) has an associated weight \(w_{ij}\). The goal of MWM is to find a pairing \(M\) (a set of edges such that no two edges share a common vertex) that maximizes the sum of the weights of the edges included in \(M\):
\[
\text{Maximize } \sum_{(u_i, v_j) \in M} w_{ij}.
\]

In the context of cluster pairing, the MWM approach can be applied to align clusters from two different clustering results. Here, we define the problem as follows:
\begin{itemize}
    \item Let $\mathcal{C}_1$ and $\mathcal{C}_2$ be two sets of clusters obtained from different clustering solutions. Each cluster in $\mathcal{C}_1$ can be paired with each cluster in $\mathcal{C}_2$.
    \item Construct a bipartite graph where one part of vertices represents the clusters from $\mathcal{C}_1$ and the other part represents the clusters from $\mathcal{C}_2$.
    \item The edge weights in this graph are defined by the entries \(a_{ij}\) of the contingency matrix, where \(a_{ij}\) represents the number of data points that are common to the \(i\)-th cluster of $\mathcal{C}_1$ and the \(j\)-th cluster of $\mathcal{C}_2$.
\end{itemize}

The objective is to find a matching that maximizes the total weight based on the contingency matrix, thereby achieving the best possible pairing between clusters from the two clustering results. This approach ensures that the pairs of clusters that have the highest number of shared data points are matched together, providing an optimal solution for evaluating and comparing clustering solutions.

This problem can be solved in \(O(|V|^2 \cdot |E|)\) time using Edmonds' Algorithm \cite{Edmonds_1965} for Maximum Weighted Matching. In our clustering problem, where the number of clusters \(|V|\) is \(N\) and the number of edges \(|E|\) is \(N^2\), the time complexity of solving the MWM problem becomes \(O(N^4)\). However, advancements in this area were made by Gabow \cite{gabow_eff}, who later introduced a more efficient implementation of the algorithm. Gabow's improvement reduces the time complexity to \(O(N^3)\), significantly optimizing the solution for this class of problems. As a result, the Maximum Weighted Matching problem, and consequently our problem, can now be solved in \(O(N^3)\) providing a more practical and computationally feasible approach.

\subsection{Stable Matching}

The stable matching problem involves creating pairings between two sets of entities, such as a group of boys and a group of girls, where every individual has preferences over the members of the opposite group. The goal is to form pairings that respect these preferences and avoid instability.

A matching is considered unstable if there exists a pair, say (Ben and Jen), where:
\begin{enumerate}
    \item Ben is not paired with Jen.
    \item Ben prefers Jen to his current match if she exists.
    \item Jen prefers Ben to her current match if he exits.
\end{enumerate}

If such a pair exists, the matching is unstable because Ben and Jen would prefer to leave their current partners and pair with each other. A stable matching ensures that no such situation can occur, meaning that no individual can improve their situation by leaving their current match for someone else.

To achieve a stable matching, the Gale-Shapley algorithm can be broken down into the following steps:

\begin{verbatim}
function stable_matching:
    Each individual on one side (e.g., the boys) is initially unmatched, and no proposals
    have been made.
    
    while no unmatched individuals can propose or improve their match.:
    
        Each unmatched individual (boy) proposes to their most preferred partner (girl) 
        who has not yet rejected them.

        Each individual on the receiving side (girl) tentatively accepts the proposal 
        from the most preferred proposer and rejects all other proposals. If a girl is 
        already tentatively matched but receives a better proposal, she will switch to
        the new proposer and reject her previous match.
\end{verbatim}

    
    
    
    

This step-by-step approach guarantees that the final matching is stable, meaning no individual has an incentive to leave their current pairing for another potential partner.

In the worst case, where every proposer is rejected by all receivers except the last one, the Gale-Shapley algorithm has a time complexity of \(O(M \cdot N)\), where \(M\) and \(N\) are the number of individuals on each side.

\subsection{Stable Matching Based Pairing (SMBP)}

To evaluate the effectiveness of different clustering methods we propose Stable Matching Based Pairing (SMBP). In SMBP, each cluster from one clustering method expresses a preference for clusters from the other method based on the contingency matrix. Specifically, a cluster will prefer the one from the other clustering that shares the most data points with. In another way, a cluster will prefer a cluster from the other clustering that has a higher $a_{ij}$ value in the contingency matrix. This concept of preference is then translated into a stable matching problem, similar to the classical Gale-Shapley algorithm.

In our setup, one group of clusters (e.g., from the first clustering method) acts as the proposers (analogous to "boys" in the traditional stable marriage problem \cite{mcVitie}), and the other group of clusters (from the second method) acts as the receivers (analogous to "girls"). Each proposer ranks the receivers based on the number of shared data points, forming a priority list. The receivers also rank the proposers similarly.

Using these priority lists, we run the stable matching algorithm to pair clusters between the two cluster sets. The resulting stable matching indicates which clusters are most similar based on their shared data points. This process provides a robust and interpretable way to evaluate the clustering methods: by examining the number of common data points within the matched pairs, we can quantify how well the two clustering methods agree on the underlying structure of the data. 

The stable matching ensures that the resulting cluster pairs are not only based on local preferences but also globally stable, meaning no two clusters outside the matched pairs would prefer each other more than their current matches. The final evaluation metric is then derived by summing the number of shared data points within these matched pairs, providing a clear and intuitive measure of clustering quality.

The Gale-Shapley algorithm used in SMBP runs with a time complexity of \(O(M \cdot N)\), where \(M\) and \(N\) represent the number of clusters from each clustering method. In the context of our experiments, where \(M = N\), the time complexity simplifies to \(O(N^2)\). Therefore, SMBP runs in \(O(N^2)\) time for evaluating clustering methods. Additionally, SMBP's compatibility with modern machine learning frameworks such as PyTorch and TensorFlow further enhances its applicability in real-world big data scenarios. Moreover, SMBP is capable of evaluating cluster groups with different numbers of clusters, making it versatile for comparing clustering methods across diverse datasets.

\subsection{Dataset Generation}
To evaluate the performance of our model across various conditions, we generate datasets with a large number of clusters in two different configurations: \textit{balanced} and \textit{unbalanced}. This approach allows us to generalize our findings and ensure robustness across different types of data distributions.

\subsubsection{Balanced Dataset}
A \textit{balanced dataset} is characterized by an equal or nearly equal distribution of samples across all clusters or categories. Such datasets are highly desirable in machine learning tasks as they prevent the model from being biased toward a particular class, which could otherwise skew predictions. Balanced datasets typically lead to more accurate and equitable performance across all classes.

We generate the balanced dataset using the following algorithm:

\begin{verbatim}
function balanced_dataset_generator(number_of_communities, number_of_rows):

    class_labels = create_sequence(1, number_of_communities) 
    
    dataset = create_empty_list()
    
    for i in range(0, number_of_rows):
    
        random_class_label = randomly_select_from_list(class_labels)
        
        append_to_list(dataset, random_class_label)
    
    return dataset
\end{verbatim}

In this method:
\begin{itemize}
    \item \textit{number\_of\_communities} defines the number of distinct clusters.
    \item \textit{number\_of\_rows} specifies the total number of samples in the dataset.
    \item A class label is randomly selected from a predefined set of labels, ensuring that each class has an equal chance of being chosen for each sample.
\end{itemize}

\subsubsection{Unbalanced Dataset}
An \textit{unbalanced dataset} exhibits significant disparity in the distribution of classes, where one or more categories contain far fewer samples than others. This scenario is common in real-world datasets and presents a challenge in machine learning, as models trained on Unbalanced data may exhibit a strong bias toward the majority class, resulting in poor predictive performance on minority classes.

To generate an unbalanced dataset, we leverage a Gaussian (normal) distribution. By controlling the mean and standard deviation, we can simulate the unequal distribution of data points across clusters. The following algorithm is used:

\begin{verbatim}
function unbalanced_dataset_generator(number_of_communities, number_of_rows, 
mean, std):

    dataset = create_empty_list()
    
    for i in range(0, number_of_rows):
    
        random_value = randomly_select_from_Gaussian_distribution(mean, std)
        
        random_int = convert_floating_point_to_integer_value(random_value)
        
        random_clipped_int = clip(random_int, 1, number_of_communities)
        
        append_to_list(dataset, random_clipped_int)
    
    return dataset
\end{verbatim}

In this method:
\begin{itemize}
    \item The mean and std define the parameters of the Gaussian distribution. Specifically, we set the mean as $(number\_of\_communities + 1)/2$ and the standard deviation as $number\_of\_communities/4$. This choice has been found to generate a satisfactory unbalanced distribution, where the majority of data points are clustered around the middle communities, with fewer points belonging to the extremes.
    \item The \textit{random\_value} is drawn from the Gaussian distribution, and then it is rounded to the nearest integer.
    \item The \textit{clip()} function ensures that the generated value falls within the valid range of cluster labels, i.e., between 1 and \textit{number\_of\_communities}. Any value below 1 is clipped to 1, and any value above \textit{number\_of\_communities} is clipped to \textit{number\_of\_communities}.
\end{itemize}

This approach allows us to simulate real-world imbalances where certain categories may be overrepresented or underrepresented. The use of the Gaussian distribution, particularly with the chosen mean and standard deviation, generates an effective balance between common and rare categories, thereby introducing practical challenges for model training and evaluation.

\subsection{Experiments}

In this section, we evaluate the performance of the proposed Stable Matching Based Pairing (SMBP) method, along with three other pairing techniques: Maximum Weighted Matching (Baseline Method), Maximum Match Measure (MMM), and Centroid Ratio (CR). The evaluation is conducted across three distinct stages, focusing on real-world datasets and synthetic datasets of varying size and complexity.

\subsubsection{Evaluation on Real Datasets}

We begin by assessing the performance of SMBP, MMM, MWM, and CR on five well-known real-world datasets: \textit{Iris} \cite{misc_iris_53}, \textit{Dry Bean} \cite{misc_dry_bean_602}, \textit{Letter Recognition} \cite{misc_letter_recognition_59}, \textit{Obesity} \cite{misc_estimation_of_obesity_levels_based_on_eating_habits_and_physical_condition__544}, and \textit{Image Segmentation} \cite{misc_image_segmentation_50}. A detailed description of these datasets can be found in Table \ref{tab:real-datasets}. For this evaluation, we consider the true labels of each dataset as the first clustering group. The second clustering group is generated in two different ways:
\begin{enumerate}
    \item Random balanced clustering.
    \item Random unbalanced clustering.
\end{enumerate}


\begin{table*}[ht]
\centering
\caption{Summary of real-world Datasets Used in the Experiments}
\label{tab:real-datasets}
\fontsize{7}{8.4}\selectfont
\begin{tabular}{|c|c|c|c|c|p{5cm}|}
\hline
\textbf{Dataset}             & \textbf{Feature Type}      & \textbf{\# Instances} & \textbf{\# Features} & \textbf{\# Communities} & \textbf{Description} \\ \hline
\textbf{Iris}                & Real                      & 150                   & 4                    & 3                       & Contains measurements of iris flowers from three species to classify them. \\ \hline
\textbf{Dry Bean}            & Integer, Real             & 13,611                & 16                   & 7                       & Consists of physical properties of different types of dry beans to classify the bean type. \\ \hline
\textbf{Letter Recognition}  & Integer                   & 20,000                & 16                   & 26                      & Contains 16-dimensional attributes derived from the pixel values of English letters, used for letter classification. \\ \hline
\textbf{Obesity Levels}      & Integer                   & 2,111                 & 8                   & 101                       & Consists of personal and lifestyle information to classify individuals based on their obesity level. \\ \hline
\textbf{Image Segmentation}  & Real                      & 210                   & 19                   & 7                       & Contains pixel-level features to segment an image into distinct regions or objects. \\ \hline
\end{tabular}
\end{table*}

To evaluate the quality of the pairings generated by each method, we use \textit{Maximum Weighted Matching (MWM)} as the optimal baseline. The weight of each pair is calculated based on the number of shared data points between the clusters in the contingency matrix. The following procedure is applied:
\begin{enumerate}
    \item We sum the weights of the cluster pairs generated by MWM, which serves as the optimal benchmark.
    \item For each method (SMBP, MMM, CR), we sum the weights of the pairs obtained and compute an \textit{accuracy} metric as:
    \[
    \text{Accuracy} = \frac{\text{Sum of weights (method)}}{\text{Sum of weights (MWM)}}
    \]
\end{enumerate}

This process is repeated for 100 iterations to account for variability in random clustering, reporting the mean accuracy, standard deviation (std) of the accuracy, and the run time of each method.

\subsubsection{Detailed Comparison of SMBP and MMM}

In the second stage, we conduct a more in-depth comparison of SMBP and MMM, using MWM as the optimal pairing method. We evaluate their performance under three different conditions:
\begin{enumerate}
    \item Both clustering groups are generated using balanced random clustering.
    \item One clustering group is balanced, and the other is unbalanced.
    \item Both clustering groups are unbalanced.
\end{enumerate}

The evaluation is conducted on a dataset of 100,000 samples distributed across 100 clusters, and the process is repeated for 50 iterations. We report the mean accuracy, std of accuracy, and run time for each method. 

Additionally, we conduct evaluations on larger datasets with a larger number of clusters, specifically:
\begin{enumerate}
    \item One iteration with 10 million samples across 500 communities.
    \item One iteration with 20 million samples across 1,000 communities.
\end{enumerate}

Detailed information on the standard deviation and mean cluster sizes for the randomly generated datasets can be found in Table \ref{tab:random-datasets-stats}.

\begin{table*}[ht]
\centering
\caption{Cluster Size Statistics for Synthetic Datasets in the Second Experiment}
\label{tab:random-datasets-stats}
\begin{tabular}{|c|c|c|c|c|c|}
\hline
\textbf{ } & \textbf{\# Samples} & \textbf{\# Communities} & \textbf{Iterations} & \textbf{Dataset 1 (mean cluster size $\pm$ std)} & \textbf{Dataset 2 (mean cluster size $\pm$ std)} \\ \hline
1                      & 100k                & 100                     & 50                  & 1000$\pm$30.871 (balanced)             & 1000$\pm$30.267 (balanced)             \\ \hline
2                      & 100k                & 100                     & 50                  & 1000$\pm$498.805 (unbalanced)          & 1000$\pm$30.267 (balanced)             \\ \hline
3                      & 100k                & 100                     & 50                  & 1000$\pm$498.106 (unbalanced)          & 1000$\pm$498.805 (unbalanced)          \\ \hline
4                      & 10M                 & 500                     & 1                   & 20000$\pm$140.441 (balanced)            & 20000$\pm$147.735 (balanced)            \\ \hline
5                      & 10M                 & 500                     & 1                   & 20000$\pm$16245.066 (unbalanced)        & 20000$\pm$147.735 (balanced)            \\ \hline
6                      & 10M                 & 500                     & 1                   & 20000$\pm$16244.066 (unbalanced)        & 20000$\pm$16245.066 (unbalanced)        \\ \hline
7                      & 20M                 & 1000                    & 1                   & 20000$\pm$144.655 (balanced)            & 20000$\pm$140.439 (balanced)            \\ \hline
8                      & 20M                 & 1000                    & 1                   & 20000$\pm$21694.067 (unbalanced)        & 20000$\pm$140.439 (balanced)            \\ \hline
9                      & 20M                 & 1000                    & 1                   & 20000$\pm$21674.655 (unbalanced)        & 20000$\pm$21694.067 (unbalanced)        \\ \hline
\end{tabular}
\end{table*}

\subsubsection{Detailed Comparison of SMBP and MMM in Large-scale Datasets With a Large Number of Clusters}

In the third stage, we evaluate the performance of SMBP and MMM in extremely large-scale scenarios. Due to the significant computational complexity of MWM in such large-scale experiments, we exclude MWM from this evaluation. Instead, we introduce a normalized accuracy metric that compares the effectiveness of SMBP and MMM:
\[
\text{Normalized Accuracy} = \frac{\text{Sum of weights (method)}}{\max(\text{Sum of weights (SMBP)}, \text{Sum of weights (MMM)})}
\]

This metric provides a relative measure of performance between SMBP and MMM without considering MWM as a baseline. Similar to our second experiment, we evaluated the performance of these models under three different conditions:

\begin{enumerate}
    \item Both clustering groups are generated using balanced random clustering.
    \item One clustering group is balanced, and the other is unbalanced.
    \item Both clustering groups are unbalanced.
\end{enumerate}

The experiments are conducted on the following large-scale datasets with a large number of clusters, for one iteration under those three conditions:

\begin{enumerate}
    \item 40 million samples distributed across 2,000 communities.
    \item 200 million samples distributed across 5,000 communities.
    \item 400 million samples distributed across 10,000 communities.
\end{enumerate}

The mean and standard deviation of the cluster sizes for the randomly generated datasets used in these experiments are provided in Table \ref{tab:random-datasets-stats2}.

\begin{table*}[ht]
\centering
\caption{Cluster Size Statistics for Synthetic Datasets in the Third Experiment}
\label{tab:random-datasets-stats2}
\begin{tabular}{|c|c|c|c|c|c|}
\hline
\textbf{ } & \textbf{\# Samples} & \textbf{\# Communities} & \textbf{Dataset 1 (mean cluster size $\pm$ std)} & \textbf{Dataset 2 (mean cluster size $\pm$ std)} \\ \hline
1                      & 40M                & 2000                                  & 20000$\pm$141.538 (balanced)             & 20000$\pm$140.073 (balanced)             \\ \hline
2                      & 40M                & 2000                               & 20000$\pm$29772.543 (unbalanced)          & 20000$\pm$140.073 (balanced)             \\ \hline
3                      & 40M                & 2000                                  & 20000$\pm$29761.746 (unbalanced)          & 20000$\pm$29772.543 (unbalanced)          \\ \hline
4                      & 200M                 & 5000                                 & 40000$\pm$200.855 (balanced)            & 40000$\pm$202.585 (balanced)            \\ \hline
5                      & 200M                 & 5000                                & 40000$\pm$92266.063 (unbalanced)        & 40000$\pm$202.585 (balanced)            \\ \hline
6                      & 200M                 & 5000                                 & 40000$\pm$92280.326 (unbalanced)        & 40000$\pm$92266.063 (unbalanced)        \\ \hline
7                      & 400M                 & 10000                                  & 40000$\pm$199.930 (balanced)            & 40000$\pm$199.307 (balanced)            \\ \hline
8                      & 400M                 & 10000                                  & 40000$\pm$129638.312 (unbalanced)        & 40000$\pm$200.203 (balanced)            \\ \hline
9                      & 400M                 & 10000                              & 40000$\pm$129623.660 (unbalanced)        & 40000$\pm$129566.294 (unbalanced)        \\ \hline
\end{tabular}
\end{table*}

\section{Results}

\subsection{Results on Real Datasets}

The performance of non-optimal methods across five real-world datasets is evaluated based on their accuracy and run-time. Notably, Stable Matching and Maximum Match Measure (MMM) often achieve highly comparable results, showcasing a competitive relationship in terms of accuracy and speed. Both methods frequently outperform Centroid Ratio, which tends to lag behind in accuracy. SMBP and MMM consistently exhibit close performance in both accuracy and run-time across multiple datasets, reflecting their competitiveness in achieving effective pairings. Detailed performance metrics for each dataset can be found in Tables~\ref{tab:iris-metrics-table} to \ref{tab:segmentation-metrics-table}.

\subsubsection{Iris Dataset}

In the Iris dataset with the balanced random dataset, SMBP achieves the highest accuracy among the non-optimal methods with \textbf{99.50\%} accuracy, closely followed by MMM with \textbf{99.15\%}. Centroid Ratio, in contrast, performs significantly worse, with \textbf{88.06\%} accuracy. In terms of run-time, MMM is the fastest with \textbf{0.0000953} seconds, slightly outperforming SMBP, which clocks in at \textbf{0.0001281} seconds (Table~\ref{tab:iris-metrics-table}).

For the Iris dataset with the unbalanced random dataset, a similar trend emerges, with SMBP recording the best non-optimal accuracy of \textbf{99.31\%}, again closely followed by MMM at \textbf{99.10\%}. Centroid Ratio remains behind at \textbf{88.17\%}. MMM also takes the lead in terms of run-time at \textbf{0.0000616} seconds, whereas SMBP finishes at \textbf{0.0000906} seconds (Table~\ref{tab:iris-metrics-table}).

\begin{table*}[ht]
\centering
\caption{Performance Metrics for Different Models on the Iris Dataset}
\label{tab:iris-metrics-table}
\begin{tabular}{|c|c|c|c|}
\hline
\textbf{Dataset Type} & \textbf{Model}                  & \textbf{Run Time in Seconds} & \textbf{Accuracy$\pm$Std} \\ \hline
\multirow{4}{*}{Balanced} 
& Maximum Weighted Matching & 0.0004030 & 1$\pm$0 \\ \cline{2-4} 
& Stable Matching Based Pairing         & 0.0001281 & \textbf{0.9950}$\pm$\textbf{0.015}  \\ \cline{2-4} 
& Centroid Ratio           & 0.0046806 & 0.8806$\pm$0.106 \\ \cline{2-4} 
& Maximum Match Measure    & \textbf{0.0000953} & 0.9915$\pm$0.023 \\ \hline

\multirow{4}{*}{Unbalanced} 
& Maximum Weighted Matching & 0.0003517 & 1$\pm$0 \\ \cline{2-4} 
& Stable Matching Based Pairing          & 0.0000906 & \textbf{0.9931}$\pm$\textbf{0.016} \\ \cline{2-4} 
& Centroid Ratio           & 0.0040929 & 0.8817$\pm$0.101 \\ \cline{2-4} 
& Maximum Match Measure    & \textbf{0.0000616} & 0.9910$\pm$0.019 \\ \hline
\end{tabular}
\end{table*}

\subsubsection{Dry Bean Dataset}

On the Dry Bean dataset with the balanced random dataset, MMM slightly edges out SMBP with \textbf{99.17\%} accuracy compared to \textbf{99.16\%} in the balanced case, with both methods being nearly identical in performance. Centroid Ratio, on the other hand, performs noticeably worse, achieving \textbf{94.53\%} accuracy. MMM again demonstrates the fastest run-time at \textbf{0.0002897} seconds, beating SMBP’s \textbf{0.0003666} seconds (Table~\ref{tab:dry-metrics-table}).

In the Dry Bean dataset with the unbalanced random dataset, SMBP achieves the highest accuracy among the non-optimal methods with \textbf{98.877\%}, closely followed by MMM with \textbf{98.876\%}. However, MMM has a faster run-time, completing in \textbf{0.0001933} seconds, ahead of SMBP’s \textbf{0.0004126} seconds. Centroid Ratio remains less competitive at \textbf{84.45\%} accuracy (Table~\ref{tab:dry-metrics-table}).

\begin{table*}[ht]
\centering
\caption{Performance Metrics for Different Models on the Dry Bean Dataset}
\label{tab:dry-metrics-table}
\begin{tabular}{|c|c|c|c|}
\hline
\textbf{Dataset Type} & \textbf{Model}                  & \textbf{Run Time in Seconds} & \textbf{Accuracy$\pm$Std} \\ \hline
\multirow{4}{*}{Balanced} 
& Maximum Weighted Matching & 0.0027854 & 1$\pm$0 \\ \cline{2-4} 
& Stable Matching Based Pairing          & 0.0003666 & 0.9916$\pm$0.00789  \\ \cline{2-4} 
& Centroid Ratio           & 0.0866872 & 0.9453$\pm$0.01934 \\ \cline{2-4} 
& Maximum Match Measure    & \textbf{0.0002897} & \textbf{0.9917}$\pm$\textbf{0.00780} \\ \hline

\multirow{4}{*}{Unbalanced} 
& Maximum Weighted Matching & 0.0027030 & 1$\pm$0 \\ \cline{2-4} 
& Stable Matching Based Pairing          & 0.0004126 & \textbf{0.98877}$\pm$\textbf{0.01012} \\ \cline{2-4} 
& Centroid Ratio           & 0.0872055 & 0.84450$\pm$0.06579 \\ \cline{2-4} 
& Maximum Match Measure    & \textbf{0.0001933} & 0.98876$\pm$0.01032 \\ \hline
\end{tabular}
\end{table*}

\subsubsection{Letter Recognition Dataset}

For the Letter Recognition dataset with the balanced random dataset, MMM and SMBP are again closely matched, with MMM slightly outperforming SMBP at \textbf{98.55\%} vs. \textbf{98.50\%} accuracy. Centroid Ratio delivers its weakest performance yet, with an accuracy of \textbf{75.47\%}. MMM runs in \textbf{0.0029094} seconds, slightly slower than SMBP’s \textbf{0.0016922} seconds (Table~\ref{tab:letter-metrics-table}).

In the Letter Recognition dataset with the unbalanced random dataset, SMBP achieves a marginally better accuracy of \textbf{98.10\%}, compared to MMM’s \textbf{98.00\%}, though MMM is still faster with a run-time of \textbf{0.0028215} seconds compared to SMBP’s \textbf{0.0043918} seconds (Table~\ref{tab:letter-metrics-table}).

\begin{table*}[ht]
\centering
\caption{Performance Metrics for Different Models on the Letter Recognition Dataset}
\label{tab:letter-metrics-table}
\begin{tabular}{|c|c|c|c|}
\hline
\textbf{Dataset Type} & \textbf{Model}                  & \textbf{Run Time in Seconds} & \textbf{Accuracy$\pm$Std} \\ \hline
\multirow{4}{*}{Balanced} 
& Maximum Weighted Matching & 0.0467686 & 1$\pm$0 \\ \cline{2-4} 
& Stable Matching Based Pairing          & \textbf{0.0016922} & 0.9850$\pm$0.00669  \\ \cline{2-4} 
& Centroid Ratio           & 0.1716876 & 0.7547$\pm$0.03533 \\ \cline{2-4} 
& Maximum Match Measure    & 0.0029094 & \textbf{0.9855}$\pm$\textbf{0.00661} \\ \hline

\multirow{4}{*}{Unbalanced} 
& Maximum Weighted Matching & 0.0799929 & 1$\pm$0 \\ \cline{2-4} 
& Stable Matching Based Pairing          & 0.0043918 & \textbf{0.9810}$\pm$0.00844 \\ \cline{2-4} 
& Centroid Ratio           & 0.1711400 & 0.7559$\pm$0.03203 \\ \cline{2-4} 
& Maximum Match Measure    & \textbf{0.0028215} & 0.9800$\pm$\textbf{0.00843} \\ \hline
\end{tabular}
\end{table*}

\subsubsection{Obesity Dataset}

In the Obesity dataset with the balanced random dataset, SMBP edges out MMM with the highest non-optimal accuracy of \textbf{93.26\%}, while MMM is close behind at \textbf{92.96\%}. Centroid Ratio performs quite poorly here, with an accuracy of \textbf{16.87\%}. In terms of run-time, SMBP completes in \textbf{0.1232256} seconds, compared to MMM’s \textbf{0.1250238} seconds (Table~\ref{tab:letter-metrics-table}).

For the Obesity dataset with the unbalanced random dataset, MMM emerges slightly ahead with \textbf{94.47\%} accuracy, compared to SMBP’s \textbf{94.10\%}. Centroid Ratio continues to perform poorly, at \textbf{16.15\%} accuracy. MMM is faster with a run-time of \textbf{0.1255490} seconds, compared to SMBP’s \textbf{0.1814925} seconds (Table~\ref{tab:obesity-metrics-table}).

\begin{table*}[ht]
\centering
\caption{Performance Metrics for Different Models on the Obesity Dataset}
\label{tab:obesity-metrics-table}
\begin{tabular}{|c|c|c|c|}
\hline
\textbf{Dataset Type} & \textbf{Model}                  & \textbf{Run Time in Seconds} & \textbf{Accuracy$\pm$Std} \\ \hline
\multirow{4}{*}{Balanced} 
& Maximum Weighted Matching & 0.1451058 & 1$\pm$0 \\ \cline{2-4} 
& Stable Matching Based Pairing          & \textbf{0.1232256} & \textbf{0.9326}$\pm$0.01597  \\ \cline{2-4} 
& Centroid Ratio           & 0.1697107 & 0.1687$\pm$0.03351 \\ \cline{2-4} 
& Maximum Match Measure    & 0.1250238 & 0.9296$\pm$\textbf{0.01491} \\ \hline

\multirow{4}{*}{Unbalanced} 
& Maximum Weighted Matching & 0.1419915 & 1$\pm$0 \\ \cline{2-4} 
& Stable Matching Based Pairing         & 0.1814925 & 0.9410$\pm$0.01592 \\ \cline{2-4} 
& Centroid Ratio           & 0.1724453 & 0.1615$\pm$0.03618 \\ \cline{2-4} 
& Maximum Match Measure    & \textbf{0.1255490} & \textbf{0.9447}$\pm$\textbf{0.01377} \\ \hline
\end{tabular}
\end{table*}

\subsubsection{Image Segmentation Dataset}

In the Image Segmentation dataset with the balanced random dataset, MMM achieves the highest accuracy among the non-optimal methods, with \textbf{98.50\%}, while SMBP records \textbf{97.99\%}. Centroid Ratio, on the other hand, is notably weaker, at \textbf{63.23\%}. MMM also shows superior speed, finishing in \textbf{0.0002063} seconds, outperforming SMBP, which completes in \textbf{0.0002769} seconds (Table~\ref{tab:segmentation-metrics-table}).

In the Image Segmentation dataset with the unbalanced random dataset, MMM and SMBP both achieve strong results, with MMM slightly outperforming SMBP at \textbf{97.26\%} versus \textbf{97.24\%}. MMM is faster as well, with a run-time of \textbf{0.0001685} seconds compared to SMBP’s \textbf{0.0002639} seconds (Table~\ref{tab:segmentation-metrics-table}).

\begin{table*}[ht]
\centering
\caption{Performance Metrics for Different Models on the Image Segmentation Dataset}
\label{tab:segmentation-metrics-table}
\begin{tabular}{|c|c|c|c|}
\hline
\textbf{Dataset Type} & \textbf{Model}                  & \textbf{Run Time in Seconds} & \textbf{Accuracy$\pm$Std} \\ \hline
\multirow{4}{*}{Balanced} 
& Maximum Weighted Matching & 0.0014673 & 1$\pm$0 \\ \cline{2-4} 
& Stable Matching Based Pairing          & 0.0002769 & 0.9799$\pm$0.02704  \\ \cline{2-4} 
& Centroid Ratio           & 0.0093185 & 0.6323$\pm$0.10732 \\ \cline{2-4} 
& Maximum Match Measure    & \textbf{0.0002063} & \textbf{0.9850}$\pm$\textbf{0.02164} \\ \hline

\multirow{4}{*}{Unbalanced} 
& Maximum Weighted Matching & 0.0017026 & 1$\pm$0 \\ \cline{2-4} 
& Stable Matching Based Pairing        & 0.0002639 & 0.9724$\pm$0.03442 \\ \cline{2-4} 
& Centroid Ratio           & 0.0087252 & 0.6502$\pm$0.11772 \\ \cline{2-4} 
& Maximum Match Measure    & \textbf{0.0001685} & \textbf{0.9726}$\pm$\textbf{0.03200} \\ \hline
\end{tabular}
\end{table*}

\subsection{Results on Detailed Comparison of SMBP and MMM}

In this section, we provide a detailed comparison of Stable Matching Based Pairing (SMBP) and Maximum Match Measure (MMM) based on performance metrics across random datasets of varying sizes. We have excluded Centroid Ratio from this experiment as it demonstrated inferior performance in both run time and accuracy compared to the other two methods in the first experiment. The results for these metrics are summarized in Tables~\ref{tab:hunred-metrics-table} to \ref{tab:twM-metrics-table}.

\subsubsection{Datasets with 100K Data Points and 100 Communities}

For the random datasets with 100K data points and 100 communities, SMBP demonstrates superior efficiency compared to MMM and Maximum Weighted Matching (MWM). SMBP achieves the best run-time of \textbf{0.014965} seconds in balanced dataset scenarios, whereas MMM takes \textbf{0.120338} seconds. In terms of accuracy, MMM leads slightly with \textbf{98.27\%} compared to SMBP’s \textbf{98.23\%}. When dealing with one balanced and one unbalanced dataset, SMBP again outperforms MMM in run-time (\textbf{0.062017} seconds vs. \textbf{0.118971} seconds) and achieves a higher accuracy of \textbf{97.55\%} compared to MMM’s \textbf{97.49\%}. For both unbalanced datasets, SMBP’s run-time of \textbf{0.052161} seconds is better than MMM’s \textbf{0.117342} seconds, although MMM achieves a slightly higher accuracy of \textbf{96.58\%} compared to SMBP’s \textbf{96.56\%} (Table~\ref{tab:hunred-metrics-table}).

\subsubsection{Datasets with 10M Data Points and 500 Communities}

In the scenario with 10M data points and 500 communities, SMBP exhibits a significant advantage in run-time efficiency, completing the task in \textbf{0.283} seconds in balanced datasets, compared to MMM’s \textbf{14.100} seconds and MWM’s \textbf{153.492} seconds. SMBP achieves the highest accuracy of \textbf{99.18\%}, closely followed by MMM with \textbf{99.14\%}. For one balanced and one unbalanced dataset, SMBP’s run-time is \textbf{5.219} seconds with an accuracy of \textbf{98.80\%}, whereas MMM takes \textbf{14.168} seconds and achieves an accuracy of \textbf{98.73\%}. In the fully unbalanced scenario, SMBP’s run-time of \textbf{6.410} seconds outperforms MMM’s \textbf{14.189} seconds, with SMBP achieving a better accuracy of \textbf{97.91\%} compared to MMM’s \textbf{97.85\%} (Table~\ref{tab:tenM-metrics-table}).

\begin{table*}[ht]
\centering
\caption{Performance Metrics for Different Models on the random datasets with 100K data points and 100 communities for 50 iterations}
\label{tab:hunred-metrics-table}
\begin{tabular}{|c|c|c|c|}
\hline
\textbf{Datasets Types} & \textbf{Model}                  & \textbf{Run Time in Seconds} & \textbf{Accuracy$\pm$Std} \\ \hline
\multirow{3}{*}{Both balanced} 
& Maximum Weighted Matching & 1.362103 & 1$\pm$0 \\ \cline{2-4} 
& Stable Matching Based Pairing          & \textbf{0.014965} & 0.9823$\pm$0.00438  \\ \cline{2-4} 
& Maximum Match Measure    & 0.120338 & \textbf{0.9827}$\pm$\textbf{0.00370} \\ \hline

\multirow{3}{*}{One balanced, and one unbalanced} 
& Maximum Weighted Matching & 2.148780 & 1$\pm$0 \\ \cline{2-4} 
& Stable Matching Based Pairing          & \textbf{0.062017} & \textbf{0.9755}$\pm$0.00521  \\ \cline{2-4} 
& Maximum Match Measure    & 0.118971 & 0.9749$\pm$\textbf{0.00503} \\ \hline

\multirow{3}{*}{Both unbalanced} 
& Maximum Weighted Matching & 2.422409 & 1$\pm$0 \\ \cline{2-4} 
& Stable Matching Based Pairing        & \textbf{0.052161} & 0.9656$\pm$0.00695 \\ \cline{2-4} 
& Maximum Match Measure    & 0.117342 & \textbf{0.9658}$\pm$\textbf{0.00591} \\ \hline
\end{tabular}
\end{table*}

\begin{table*}[ht]
\centering
\caption{Performance Metrics for Different Models on the random datasets with 10M data points and 500 communities for 1 iteration}
\label{tab:tenM-metrics-table}
\begin{tabular}{|c|c|c|c|}
\hline
\textbf{Datasets Types} & \textbf{Model}                  & \textbf{Run Time in Seconds} & \textbf{Accuracy} \\ \hline
\multirow{3}{*}{Both balanced} 
& Maximum Weighted Matching & 153.492 & 1 \\ \cline{2-4} 
& Stable Matching Based Pairing          & \textbf{0.283} & \textbf{0.9918}  \\ \cline{2-4} 
& Maximum Match Measure    & 14.100 & 0.9914 \\ \hline

\multirow{3}{*}{One balanced, and one unbalanced} 
& Maximum Weighted Matching & 226.018 & 1 \\ \cline{2-4} 
& Stable Matching Based Pairing          & \textbf{5.219} & \textbf{0.9880}  \\ \cline{2-4} 
& Maximum Match Measure    & 14.168 & 0.9873 \\ \hline

\multirow{3}{*}{Both unbalanced} 
& Maximum Weighted Matching & 257.529 & 1 \\ \cline{2-4} 
& Stable Matching Based Pairing          & \textbf{6.410} & \textbf{0.9791}  \\ \cline{2-4} 
& Maximum Match Measure    & 14.189 & 0.9785 \\ \hline
\end{tabular}
\end{table*}

\subsubsection{Datasets with 20M Data Points and 1000 Communities}

For the largest datasets with 20M data points and 1000 communities, SMBP maintains its efficiency with a run-time of \textbf{1.13} seconds in balanced datasets, significantly outperforming MMM’s \textbf{117.90} seconds and MWM’s \textbf{1162.07} seconds. Although MMM achieves the highest accuracy of \textbf{99.00\%}, SMBP follows closely with \textbf{98.96\%}. In the case of one balanced and one unbalanced dataset, SMBP’s run-time of \textbf{27.53} seconds is more efficient than MMM’s \textbf{117.66} seconds, with SMBP’s accuracy of \textbf{98.60\%} just behind MMM’s \textbf{98.71\%}. For both unbalanced datasets, SMBP’s run-time of \textbf{21.23} seconds outperforms MMM’s \textbf{117.52} seconds, while MMM achieves a slightly better accuracy of \textbf{98.24\%} compared to SMBP’s \textbf{98.20\%} (Table~\ref{tab:twM-metrics-table}).

\begin{table*}[ht]
\centering
\caption{Performance Metrics for Different Models on the random datasets with 20M data points and 1000 communities for 1 iteration}
\label{tab:twM-metrics-table}
\begin{tabular}{|c|c|c|c|}
\hline
\textbf{Datasets Types} & \textbf{Model}                  & \textbf{Run Time in Seconds} & \textbf{Accuracy} \\ \hline
\multirow{3}{*}{Both balanced} 
& Maximum Weighted Matching & 1162.07 & 1 \\ \cline{2-4} 
& Stable Matching Based Pairing          & \textbf{1.13} & 0.9896 \\ \cline{2-4} 
& Maximum Match Measure    & 117.90 & \textbf{0.9900} \\ \hline

\multirow{3}{*}{One balanced, and one unbalanced} 
& Maximum Weighted Matching & 1341.89 & 1 \\ \cline{2-4} 
& Stable Matching Based Pairing          & \textbf{27.53} & 0.9860 \\ \cline{2-4} 
& Maximum Match Measure    & 117.66 & \textbf{0.9871} \\ \hline

\multirow{3}{*}{Both unbalanced} 
& Maximum Weighted Matching & 1466.43 & 1 \\ \cline{2-4} 
& Stable Matching Based Pairing          & \textbf{21.23} & 0.9820 \\ \cline{2-4} 
& Maximum Match Measure    & 117.52 & \textbf{0.9824} \\ \hline
\end{tabular}
\end{table*}

Overall, the results demonstrate that Stable Matching Based Pairing (SMBP) significantly outperforms Maximum Match Measure (MMM) in run-time efficiency, particularly on datasets with a larger number of clusters. However, the accuracy of both methods remains competitive, with close results observed across different dataset sizes.

\subsection{Results on Comparison of SMBP and MMM on Datasets with a large number of clusters}

The performance of the SMBP and MMM models was rigorously evaluated on large-scale datasets with a large number of clusters, to highlight their scalability and computational efficiency. Three different configurations were tested, where the datasets consisted of increasing sizes and community structures, providing insight into how each model handles big data.


For datasets with 40 million data points and 2000 communities, SMBP exhibits exceptional performance. In scenarios where both datasets are balanced, SMBP completes the task in \textbf{5.0} seconds, which is markedly faster than MMM, which takes \textbf{960.9} seconds. Both models achieve high accuracy, with SMBP attaining a normalized accuracy of \textbf{100\%} compared to MMM's \textbf{99.98\%}. When dealing with one balanced and one unbalanced dataset, SMBP's run-time is \textbf{126.6} seconds, while MMM requires \textbf{956.2} seconds. SMBP’s normalized accuracy in this scenario is \textbf{99.96\%}, slightly below MMM's \textbf{100\%}. For both datasets being unbalanced, SMBP completes the task in \textbf{66.2} seconds, significantly outpacing MMM’s \textbf{952.5} seconds. Here, SMBP’s accuracy is \textbf{99.95\%}, compared to MMM’s \textbf{100\%} (Table~\ref{tab:two-metrics-table}).

\begin{table*}[ht]
\centering
\caption{Performance Metrics for Different Models on the random datasets with 40M data points and 2000 communities for 1 iteration}
\label{tab:two-metrics-table}
\begin{tabular}{|c|c|c|c|}
\hline
\textbf{Datasets Types} & \textbf{Model}                  & \textbf{Run Time in Seconds} & \textbf{Normalized Accuracy} \\ \hline
\multirow{2}{*}{Both balanced} 
& Stable Matching Based Pairing          & \textbf{5.0} & \textbf{1}  \\ \cline{2-4} 
& Maximum Match Measure    & 960.9 & 0.9998 \\ \hline

\multirow{2}{*}{One balanced, and one unbalanced} 
& Stable Matching Based Pairing          & \textbf{126.6} & 0.9996  \\ \cline{2-4} 
& Maximum Match Measure    & 956.2 & \textbf{1} \\ \hline

\multirow{2}{*}{Both unbalanced} 
& Stable Matching Based Pairing          & \textbf{66.2} & \textbf{0.9995}  \\ \cline{2-4} 
& Maximum Match Measure    & 952.5 & \textbf{1} \\ \hline
\end{tabular}
\end{table*}

In the case of datasets with 200 million data points and 5000 communities, SMBP continues to show superior performance. For both balanced datasets, SMBP completes the task in \textbf{34.2} seconds, whereas MMM does not finish within the 7200-second limit. In scenarios involving one balanced and one unbalanced dataset, SMBP's run-time is \textbf{1252.2} seconds, again outpacing MMM, which MMM does not finish within the 7200-second limit. For both datasets being unbalanced, SMBP completes the task in \textbf{548.1} seconds, while MMM does not finish within the 7200-second limit (Table~\ref{tab:three-metrics-table}).

\begin{table*}[ht]
\centering
\caption{Performance Metrics for Different Models on the random datasets with 200M data points and 5000 communities for 1 iteration}
\label{tab:three-metrics-table}
\begin{tabular}{|c|c|c|c|}
\hline
\textbf{Datasets Types} & \textbf{Model}                  & \textbf{Run Time in Seconds} \\ \hline
\multirow{2}{*}{Both balanced} 
& Stable Matching Based Pairing          & \textbf{34.2}  \\ \cline{2-3} 
& Maximum Match Measure    & Do not Finish After 7200 Seconds\\ \hline

\multirow{2}{*}{One balanced, and one unbalanced} 
& Stable Matching Based Pairing          & \textbf{1252.2}  \\ \cline{2-3} 
& Maximum Match Measure    & Do not Finish After 7200 Seconds\\ \hline

\multirow{2}{*}{Both unbalanced} 
& Stable Matching Based Pairing          & \textbf{548.1}  \\ \cline{2-3} 
& Maximum Match Measure    & Do not Finish After 7200 Seconds\\ \hline
\end{tabular}
\end{table*}

With the largest datasets with the largest number of clusters, consisting of 400 million data points and 10000 communities, SMBP still maintains a strong performance edge. On balanced datasets, SMBP finishes in \textbf{167} seconds, while MMM does not complete the task within the 7200-second limit. For one balanced and one unbalanced dataset, SMBP's run-time is \textbf{5010} seconds, with MMM failing to finish in 7200-second limit. When both datasets are unbalanced, SMBP completes the task in \textbf{1771} seconds, whereas MMM does not finish within the time frame (Table~\ref{tab:four-metrics-table}).

\begin{table*}[ht]
\centering
\caption{Performance Metrics for Different Models on the random datasets with 400M data points and 10000 communities for 1 iteration}
\label{tab:four-metrics-table}
\begin{tabular}{|c|c|c|c|}
\hline
\textbf{Datasets Types} & \textbf{Model}                  & \textbf{Run Time in Seconds} \\ \hline
\multirow{2}{*}{Both balanced} 
& Stable Matching Based Pairing          & \textbf{167}  \\ \cline{2-3} 
& Maximum Match Measure    & Do not Finish After 7200 Seconds\\ \hline

\multirow{2}{*}{One balanced, and one unbalanced} 
& Stable Matching Based Pairing          & \textbf{5010}  \\ \cline{2-3} 
& Maximum Match Measure    & Do not Finish After 7200 Seconds\\ \hline

\multirow{2}{*}{Both unbalanced} 
& Stable Matching Based Pairing          & \textbf{1771}  \\ \cline{2-3} 
& Maximum Match Measure    & Do not Finish After 7200 Seconds\\ \hline
\end{tabular}
\end{table*}

Overall, SMBP is clearly superior in handling datasets with a large number of clusters, showing a major computational advantage without compromising accuracy. This positions SMBP as a highly efficient and effective method for large-scale community detection in data with millions of points.

\section{Discussion}

One of the potential concerns with the Gale-Shapley algorithm used in Stable Matching Based Pairing (SMBP), is the issue of proposer-optimality \cite{Nisan_Roughgarden_Tardos_Vazirani_2007}. In other words, the stable matching produced by the Gale-Shapley algorithm is typically optimal for the proposing side but not necessarily for the receivers. This could raise concerns that SMBP might be biased towards the clusters that act as proposers, potentially leading to suboptimal results for the receiving clusters. However, Anna Karlin and Yuval Peres \cite{karlin2017game} provide important insights that mitigate this concern. They demonstrated that if all entries in each row are distinct and also if all entries in each column of the contingency matrix are distinct, the stable matching solution is unique. This implies that, under these conditions, it does not matter which group acts as the proposer—the resulting stable matching will be the same for both sides, thereby ensuring fairness and eliminating any bias towards the proposing group. In the context of SMBP, this result is highly relevant. Given that we are working with large-scale datasets with a large number of clusters, it is extremely rare for two entries in the same row or column of the contingency matrix to have identical values. The high probability of uniqueness of these values means that the stable matching solution is almost always effectively unbiased in big data, as the distinct entries ensure the matching is independent of which group is the proposer. As a result, in the case of big data, we can confidently state that SMBP does not suffer from the proposer-optimality problem typically associated with the Gale-Shapley algorithm.

Another concern that may arise regarding SMBP is its applicability to datasets with a large number of communities, such as more than 1000 clusters. Some may argue that datasets of this scale are currently unrealistic, and it could take many years before such datasets become widely available. However, it’s important to note that datasets of this size already exist, and their number is gradually increasing. For example, the ImageNet dataset \cite{yang2019fairer} contains thousands of communities, demonstrating that datasets of this scale are not as rare or strange as some might believe. Although these large-scale datasets with a large number of clusters, are not yet the norm, they represent the future of big data, especially in fields such as computer vision and natural language processing. As of now, community detection in these large datasets has not posed significant problems. Current methods can handle evaluations efficiently, even for datasets with thousands of clusters, without users needing to spend excessive amounts of time evaluating clustering accuracy. However, as we move into an era where datasets will likely grow much larger, containing tens or hundreds of thousands of communities, traditional methods will struggle to keep up with the computational demands. In such cases, SMBP, with its $O(N^2)$ time complexity, presents itself as a scalable and robust solution. It will be well-suited for handling the complexities of these future datasets, where conventional approaches may no longer suffice.

\section{Conclusion}

In this paper, we introduced the Stable Matching Based Pairing (SMBP) algorithm as an efficient external validity index for clustering comparisons, specifically designed for large-scale datasets with a large number of clusters. By leveraging the stable matching framework, SMBP provides a significant improvement in computational efficiency while maintaining high accuracy when compared to traditional methods like Maximum Weighted Matching (MWM) and Maximum Match Measure (MMM). We conducted experiments on both real-world datasets and balanced as well as unbalanced synthetic datasets, with the results demonstrating its efficiency across all of them, making it particularly suitable for handling large-scale datasets with a large number of clusters.

The results clearly indicate that SMBP is not only computationally efficient—thanks to its $O(N^2)$ time complexity-but also effective in accurately pairing clusters. Its compatibility with modern machine learning frameworks such as PyTorch and TensorFlow further enhances its applicability in real-world big data scenarios. Moreover, SMBP is capable of evaluating cluster groups with different numbers of clusters, making it versatile for comparing clustering methods across diverse datasets. Future work could explore optimizing SMBP for even datasets with a larger number of clusters and applying the algorithm to other domains of community detection and clustering evaluation, ensuring its place as a robust and scalable solution in the evolving landscape of data science.







\bibliographystyle{cas-model2-names}

\bibliography{main}

\bio{figs/my_picture2} 
\newline
\textbf{Mohammad Yasin Karbasian} received his Bachelor's degree in Computer Science from Isfahan University of Technology (IUT) in Iran. His academic and research pursuits are centered around data science, machine learning, deep learning, artificial intelligence, Graph Neural Networks, and Natural Language Processing. His research is currently progressing towards publication in esteemed peer-reviewed journals.
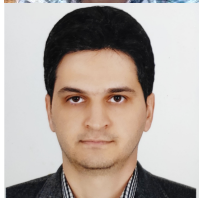
\endbio

\bio{figs/raminjavadi}
\newline
\textbf{Ramin Javadi} received his PhD on combinatorics from Sharif University of Technology. He is currently an associate professor of mathematics at Isfahan University of Technology. His research interests touch on graph theory, combinatorics, extremal combinatorics, algebraic graph theory, geometric analysis on graphs, algorithms, computational complexity, parameterized complexity and their interconnections.  
\endbio

\end{document}